\documentclass[aps,prc,twocolumn,superscriptaddress]{revtex4}
\usepackage{graphicx,rotating,array,amsmath,amssymb,pifont}
\usepackage{dcolumn}% Align table columns on decimal point
\begin{document}
%\preprint{}
\title{Single Proton Knock-Out Reactions from $^{24,25,26}$F}
% Authors and Affiliations
%*************************************************************************************
\author{M.~Thoennessen}
  \affiliation{National Superconducting Cyclotron Laboratory,
               Michigan State University, East Lansing, MI, 48824}
  \affiliation{Michigan State University, Department of Physics and Astronomy,
               East Lansing, MI 48824}
\author{T.~Baumann}
  \affiliation{National Superconducting Cyclotron Laboratory,
               Michigan State University, East Lansing, MI, 48824}
\author{B.~A.~Brown}
  \affiliation{National Superconducting Cyclotron Laboratory,
               Michigan State University, East Lansing, MI, 48824}
  \affiliation{Michigan State University, Department of Physics and Astronomy,
               East Lansing, MI 48824}
\author{J.~Enders}
  \altaffiliation[Present address: ]{Institut f\"ur Kernphysik, Technische 
               Universit\"at Darmstadt, D-64289 Darmstadt, Germany}
  \affiliation{National Superconducting Cyclotron Laboratory,
               Michigan State University, East Lansing, MI, 48824}
\author{N.~Frank}
  \affiliation{National Superconducting Cyclotron Laboratory,
               Michigan State University, East Lansing, MI, 48824}
  \affiliation{Michigan State University, Department of Physics and Astronomy,
               East Lansing, MI 48824}
\author{P.~G.~Hansen}
  \affiliation{National Superconducting Cyclotron Laboratory,
               Michigan State University, East Lansing, MI, 48824}
  \affiliation{Michigan State University, Department of Physics and Astronomy,
               East Lansing, MI 48824}
\author{P.~Heckman}
  \affiliation{National Superconducting Cyclotron Laboratory,
               Michigan State University, East Lansing, MI, 48824}
  \affiliation{Michigan State University, Department of Physics and Astronomy,
               East Lansing, MI 48824}
\author{B.~A.~Luther}
   \altaffiliation[On leave from: ]{Department of Physics, Concordia College, 
               Moorhead, MN 56562}
  \affiliation{National Superconducting Cyclotron Laboratory,
               Michigan State University, East Lansing, MI, 48824}
\author{J.~Seitz}
  \affiliation{National Superconducting Cyclotron Laboratory,
               Michigan State University, East Lansing, MI, 48824}
  \affiliation{Michigan State University, Department of Physics and Astronomy,
               East Lansing, MI 48824}
\author{A.~Stolz}
  \affiliation{National Superconducting Cyclotron Laboratory,
               Michigan State University, East Lansing, MI, 48824}
\author{E.~Tryggestad}
   \altaffiliation[Present address: ]{Institut de Physique Nucl\'{e}aire,
                  IN$_{2}$P$_{3}$-CNRS, 91406 Orsay Cedex, France}
  \affiliation{National Superconducting Cyclotron Laboratory,
               Michigan State University, East Lansing, MI, 48824}
\email[]{thoennessen@nscl.msu.edu}
%\homepage[]{Your web page}
\date{\today}

% Abstract
%*************************************************************************************
\begin{abstract}
The cross sections of the single proton knock-out reactions from $^{24}$F, $^{25}$F, and $^{26}$F on a $^{12}$C target were measured at energies of about 50 MeV/nucleon. Ground state populations of 6.6$\pm0.9$~mb, 3.8$\pm$0.6~mb for the reactions $^{12}$C($^{24}$F,$^{23}$O) and $^{12}$C($^{25}$F,$^{24}$O) were extracted, respectively. The data were compared to calculations based on the many-body shell model and the eikonal theory. In the reaction $^{12}$C($^{26}$F,$^{25}$O) the particle instability of $^{25}$O was confirmed.

\end{abstract}
%*************************************************************************************
\maketitle
% 
%*************************************************************************************
\section{Introduction}

The structure of neutron-rich nuclei in the $p/sd$-shell region has attracted significant interest recently \cite{Bro01}. The emergence of the $N$ = 16 shell and the simultaneous disappearance of the $N$ = 20 shell sparked the interest especially in the oxygen and fluorine isotopes \cite{Oza00,Uts01,Ots01}. Interaction-, reaction-, and charge-changing cross section \cite{Oza01} as well as fragmentation cross sections \cite{Lei02} have been measured and analyzed. However, a more detailed knowledge of properties of nuclei in this region is essential to understand the nuclear structure along the dripline. 

In view of the emergence of the $N$ = 16 shell it has been suggested that the normal level structure is already influenced in $^{23}$O \cite{Kan02}. One would expect that the ground state of $^{23}$O is an $s_{1/2}$ level with strong single particle character. However, in order to explain the measured momentum distribution it was necessary to modify the core ($^{22}$O) structure  \cite{Kan02}. This interpretation is controversial because an analysis of the data based on the many-body shell model and eikonal theory did not confirm the need for any changes of the level structure in $^{23}$O \cite{Bro03,Kan03}. 

Recently, it has been shown that one-nucleon knock-out reactions are a useful tool to extract spectroscopic factors \cite{Mad01,Han01,Bro02}. The single proton stripping reactions ($^{24}$F,$^{23}$O) and ($^{25}$F,$^{24}$O) should be a sensitive tool to study the ground state properties of $^{23}$O and $^{24}$O. Although experimentally not confirmed the ground state of $^{24}$F is most likely a $3^+$ state composed of $\pi d_{5/2}$ coupled to a $\nu s_{1/2}$ \cite{Ree99} and the ground-state of $^{25}$F is supposed to be the $\pi d_{5/2}$ state. Thus the reactions ($^{24}$F,$^{23}$O) and ($^{25}$F,$^{24}$O) should have strong couplings to the ground state of $^{23}$O and $^{24}$O, respectively. 

We have measured single-proton knock-out cross section of $^{24}$F, $^{25}$F, and $^{26}$F to populate states of $^{22}$O, $^{23}$O, and $^{24}$O. In most cases it its necessary to measure the reaction in coincidence with $\gamma$-rays to cleanly identify the populated state \cite{Nav98,Han01,Baz03}. However, the search for excited states in the last bound oxygen isotopes $^{23}$O and $^{24}$O have been negative \cite{Sta02} so $\gamma$-ray coincidence measurements are not necessary in the present cases to identify the final state. The knock-out reaction of the heaviest fluorine isotope feasible up-to-date ($^{26}$F,$^{25}$O) leads to the particle-unstable nucleus $^{25}$O \cite{Lan85}.

\section{Experimental Setup}

\begin{figure}[b]
  \includegraphics[width=0.6\linewidth]{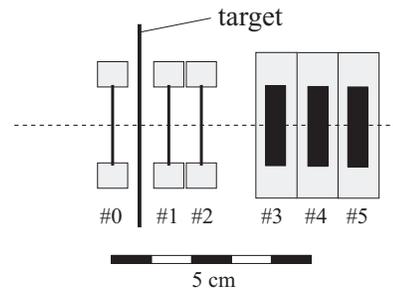}
  \caption{Experimental setup in the A1900 focal plane \cite{End03}. In the present experiment the secondary fluorine beams as well as the oxygen reaction products stopped in the first thick detector (\#3).}
  \label{f:setup}
\end{figure}

The experiments were performed at the Coupled Cyclotron Facility of the National Superconducting Cyclotron Laboratory at Michigan State University \cite{Mar01,Bau01,Mor03}. A 110 MeV/nucleon $^{48}$Ca$^{19+}$ beam was fragmented on a beryllium target. The A1900 fragment separator contained an acrylic wedge degrader at the intermediate image. The momentum acceptance of the A1900 was limited to 1\%. Details of the production of the secondary beams of $^{24}$F, $^{25}$F and $^{26}$F are listed in Table \ref{t:a1900}.

\begin{table*}
\caption{Parameters for the three secondary fluorine beams: Average beam current ($<I>$), beamtime on target ($T$), primary target (beryllium) and wedge (acrylic) thickness, magnetic rigidity of the first ($B\rho_{1,2}$) and second ($B\rho_{3,4}$) set of the A1900 dipoles, and the energy, intensity ($I$($^x$F)) and purity of the secondary fluorine isotopes. }
\begin{ruledtabular}
\begin{tabular}{rrrrrrrrrr}
 & $<I>$ & $T$ & Target & Wedge & $B\rho_{1,2}$ & $B\rho_{3,4}$ & Energy & $I$($^x$F) & Purity\\ 
& (enA) & (min) & (mg/cm$^2$) & (mg/cm$^2$) & (Tm) & (Tm) & (MeV/nucleon) & (pps/pnA) & \% \\ \hline
$^{24}$F &  20 &  610 & 587 &  971 & 3.6017 & 2.8442 & 46.7 &  82\hspace*{0.395cm}  & 24 \\
$^{25}$F & 114 & 1340 & 587 &  971 & 3.7796 & 3.0511 & 50.4 & 6.7\hspace*{0.16cm}    & 24 \\
$^{26}$F & 220 & 1590 & 376 & 1278 & 4.1451 & 3.2565 & 53.8 & 0.64 & 43 \\
\end{tabular}
\end{ruledtabular}
\label{t:a1900}
\end{table*}

At the final focus of the separator a stack of silicon detectors was located as shown in Figure \ref{f:setup}. 
The stack consisted of three 500~$\mu$m thick silicon surface barrier detectors followed by three 5000~$\mu$m thick lithium-drifted silicon-diodes. A secondary 146 mg/cm$^2$ thick $^{12}$C target was located following the first silicon detector (\#0). The setup was equivalent to that used in the measurement of proton knock-out reactions of $^8$B and $^9$C \cite{End03}. The fluorine and oxygen isotopes were stopped in the first thick silicon detector (\#3). 

The incoming fluorine isotopes were uniquely identified by the energy loss in the first detector (\#0) and the time of flight measured with respect to the cyclotron RF.

The oxygen isotopes were identified by their $\Delta E-E$ signals in the second (\#2) and third (\#3) detector after the carbon target with the additional condition that an energy loss corresponding to oxygen isotopes was deposited in the first $\Delta E$ detector (\#1) following the target. A mass identification spectrum was generated according to \cite{Shi79}:
$$\text{Mass ID}  \sim \ln (\alpha \Delta E) + (\alpha-1)\ln (E+ c\Delta E) - \alpha \ln (300) $$
where $\alpha = a-b\Delta E/T$ and $T$ is the $\Delta E$ detector thickness in ~$\mu$m. The identification spectrum was calibrated with the small percentage of oxygen isotopes which were directly produced in the A1900 production target and transmitted to the focal plane. These isotopes did not contribute to a potential background for the actual knock-out data, because of the requirement that the knock-out events had to be cleanly identified as incoming fluorine isotope in the first silicon detector (\#0).
The values of the parameters $a = 1.7715$, $b = 0.2$, and $c = 0.5$, were within the range suggested in Ref. \cite{Shi79}. The parameter $a$ was the most sensitive to a clean mass separation with deviations of less than 0.5\% resulting in significant improvements. This analysis method had previously already been applied to single proton knock-out reactions for neutron-rich carbon isotopes \cite{Kry93}.

\section{Data Analysis and Results}

\begin{figure}
  \includegraphics[width=0.9\linewidth]{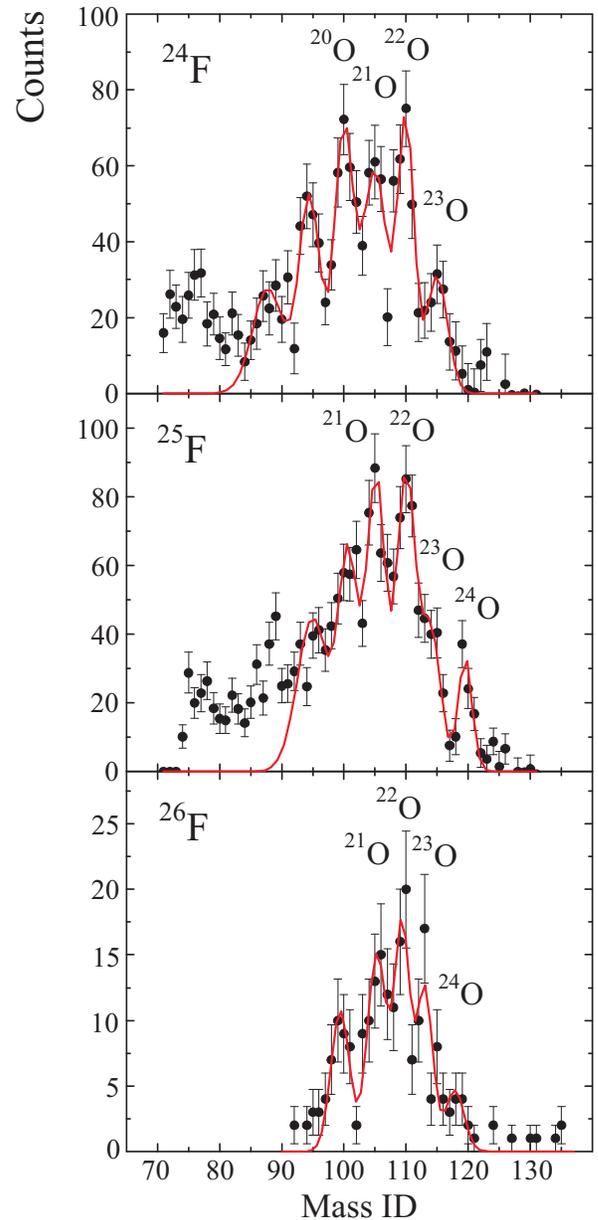}
  \caption{Oxygen isotopes populated in proton knock-out reactions from $^{24}$F (top), $^{25}$F (center) and $^{26}$F (bottom). The Mass ID parameter can be converted to the mass number by A = (Mass ID)/5, see text. (Color online)}
\label{f:data}
\end{figure}

Figure \ref{f:data} shows the populated oxygen isotopes following breakup reactions from $^{24}$F (top), $^{25}$F (center) and $^{26}$F (bottom). The Mass ID parameter was normalized according to A = (Mass ID)/5. The small overall energy dependence of the calibration results in a small shift for the $^{26}$F data so that mass 24 occurs in channel 118 instead of channel 120. In addition to the one-proton knock-out to the ground-state, which is guaranteed because no bound excited states exist in these isotopes, several oxygen isotopes of lower mass are populated. These isotopes can be either populated from sequential decay of excited states from the one-proton stripping reaction or directly from more dissipative reactions (i.e., 1p1n or deuteron, 1p2n or triton, etc.). In the following we limit the analysis to the oxygen isotopes 20$-$24. Although the structures in the mass identification spectrum indicate the population of lighter oxygen isotopes the production mechanism of these isotopes is even more uncertain.
No events of $^{25}$O were observed in the reaction of incident $^{26}$F confirming that $^{25}$O is unbound.

\begin{table}
\caption{Cross sections for the population of oxygen isotopes following breakup reactions.}
\begin{ruledtabular}
\begin{tabular}{rrrr}
 X& ($^{24}$F,X) & ($^{25}$F,X) & ($^{26}$F,X) \\ 
 & (mb) & (mb) & (mb)\\ 
\hline
$^{24}$O & ---          &  3.8$\pm$0.6 &  4.1$\pm$1.4 \\
$^{23}$O &  6.6$\pm1.0$ &  6.4$\pm$0.9 &  8.9$\pm$2.4 \\
$^{22}$O & 11.6$\pm1.6$ & 13.1$\pm$1.5 & 12.4$\pm$2.9 \\
$^{21}$O & 15.1$\pm1.7$ & 13.0$\pm$1.4 & 13.1$\pm$3.0 \\
$^{20}$O & 13.0$\pm1.9$ &  8.9$\pm$1.3 &  9.4$\pm$2.2 \\
\end{tabular}
\end{ruledtabular}
\label{t:cross}
\end{table}

The solid lines in the mass identification spectra of Figure \ref{f:data} correspond to the sum of Gaussian fits to the individual masses. The cross sections for populating these masses can then be extracted from the area of the fits, the target thickness and the number of incoming beam particles. The number of incoming particles is determined from the events in the silicon stack detectors. Any losses due to channeling and other incomplete charge collection effects in the detectors should be similar for the fluorine and the oxygen isotopes and thus cancel. This was a small effect; the difference between the number of incoming fluorine isotopes measured in the silicon stack and the first
$\Delta$-E/ToF spectrum was $<$4\%. No target-out runs were performed so that the contributions from reactions in the detectors before (\#0) and after the target (\#1) were estimated from the acceptances of the $\Delta E$ gates set in the analysis. Incoming particles were identified as F isotopes event-by-event by their $\Delta E$ and ToF gate as mentioned in section II. The acceptance of the $\Delta E$ gate was set so that isotopes reacting in the first 300$\mu m$ of detector \#0 were rejected. Oxygen isotopes produced in the last 200$\mu m$ could not be discriminated against. Similarly, fluorine isotopes reacting in the first 200$\mu m$ of detector \#1 could not be excluded from the oxygen identification condition. The reaction cross sections for single-proton knock-out in silicon were calculated with the model described in the following subsection. The contributions from the reaction in the silicon resulted then in a correction of 30$\pm 5$\%. This is in qualitative agreement with the measured target out contributions measured for the proton knock-out reactions on $^8$B and $^9$C \cite{End03}. 

The final cross sections for the population of oxygen isotopes following the breakup reactions are listed in Table \ref{t:cross}. The uncertainties are determined from the Gaussian fit with the area and the width as free parameters. In addition, the uncertainty of the contribution from the background from reactions in the detectors is added in quadrature. 

\subsection{Direct One-Proton Knock-Out Reactions to $^{23,24,25}$O}

The dominant configuration of the valence proton in the projectiles of $^{24,25,26}$F is $\pi d_{5/2}$ according to the shell model and the actual spectroscopic factors to the ground states calculated in the universal $sd$ (USD) interaction \cite{Bro88}, given in Table \ref{t:sp}, are very close to unity. The listed single-particle cross sections for the removal of this $d_{5/2}$ proton were calculated in the eikonal theory \cite{Mad01,Tos99,Han03}. Assuming that the nuclei $^{23}$O and $^{24}$O have no bound excited levels \cite{Sta02}, the measured inclusive cross section should correspond to the product of the single proton knock-out to the ground state and the spectroscopic factor. The single particle ($\sigma_{sp}$) and the calculated single proton knock-out cross sections ($\sigma_{calc}$) are compared to the experimental values ($\sigma_{exp}$) in Table \ref{t:sp}.

For this calculation the proton separation energies were taken to be 14.4, 15.0 and 16.1 MeV for $^{24,25,26}$F, respectively. The {\em rms} matter radii were deduced from experimental interaction cross-sections \cite{Oza01} as 3.20(4) ($^{23}$O) and 3.19(13) fm ($^{24}$O) for the core and 2.32 fm for the $^{12}$C target. An uncertainty of 0.1 fm in the radii translates into an uncertainty of 0.9 mb for the single-particle cross-section. The Woods-Saxon parameters for the calculation of the wave function for the relative proton-core motion were taken as the ``standard set'' of ref. \cite{Bro02}. This work deals with neutron and proton states with separation energies up to 19 MeV and strongly suggests that the eikonal calculation furnishes absolute spectroscopic factors, {\em i.e.} relating to physical occupancies of the 
shell-model states. These, however, turned out to be systematically reduced by a factor 0.5$-$0.6 relative to a many-body shell model based on effective interactions \cite{Bro02,Han03}. This effect has been attributed to contributions from the nucleon-nucleon interaction not contained in the standard shell model. Viewed in this light, and with experimental errors and theoretical uncertainties taken into account, the resulting theoretical cross sections agree well with the measured values.

\begin{table}
\caption{Spectroscopic factors ($C^2S$), single particle ($\sigma_{sp}$), calculated ($\sigma_{calc}$), and experimental ($\sigma_{exp}$) single proton knock-out cross sections. The cross sections are given in mb.}
\begin{ruledtabular}
\begin{tabular}{lrrrr}
Reaction & $C^2S$ & $\sigma_{sp}$ & $\sigma_{calc}$ & $\sigma_{exp}$ \\ \hline
($^{24}$F,$^{23}$O) & 0.91 & 8.3 & 7.6 &  6.6$\pm0.9$ \\ \hline
($^{25}$F,$^{24}$O) & 0.96 & 8.1 & 7.8 &  3.8$\pm0.6$  \\ \hline
($^{26}$F,$^{25}$O) & 0.98 & 7.8 & 7.6 &  $<$4.1$\pm1.4$ \\ 
\end{tabular}
\end{ruledtabular}
\label{t:sp}
\end{table}

The strong population of the $^{23}$O ground state in the ($^{24}$F,$^{23}$O) reaction yields additional evidence that the usual shell-model calculations give a good description of nuclei in this region \cite{Bro03}. In fact, even the simplest picture, that of the ground state of $^{23}$O described as a $[0d^6_{5/2}] \otimes 1s_{1/2}$ configuration, gives a near-quantitative description, both of the present data and of the $^{23}$O neutron knock-out reactions \cite{Kan02,Sau00}. The spin of $^{24}$F, although most likely 
3{$^-$}, is not essential for the argument, given that the proton in any case will be in the $0d_{5/2}$ state.

\begin{figure}
  \includegraphics[width=0.95\linewidth]{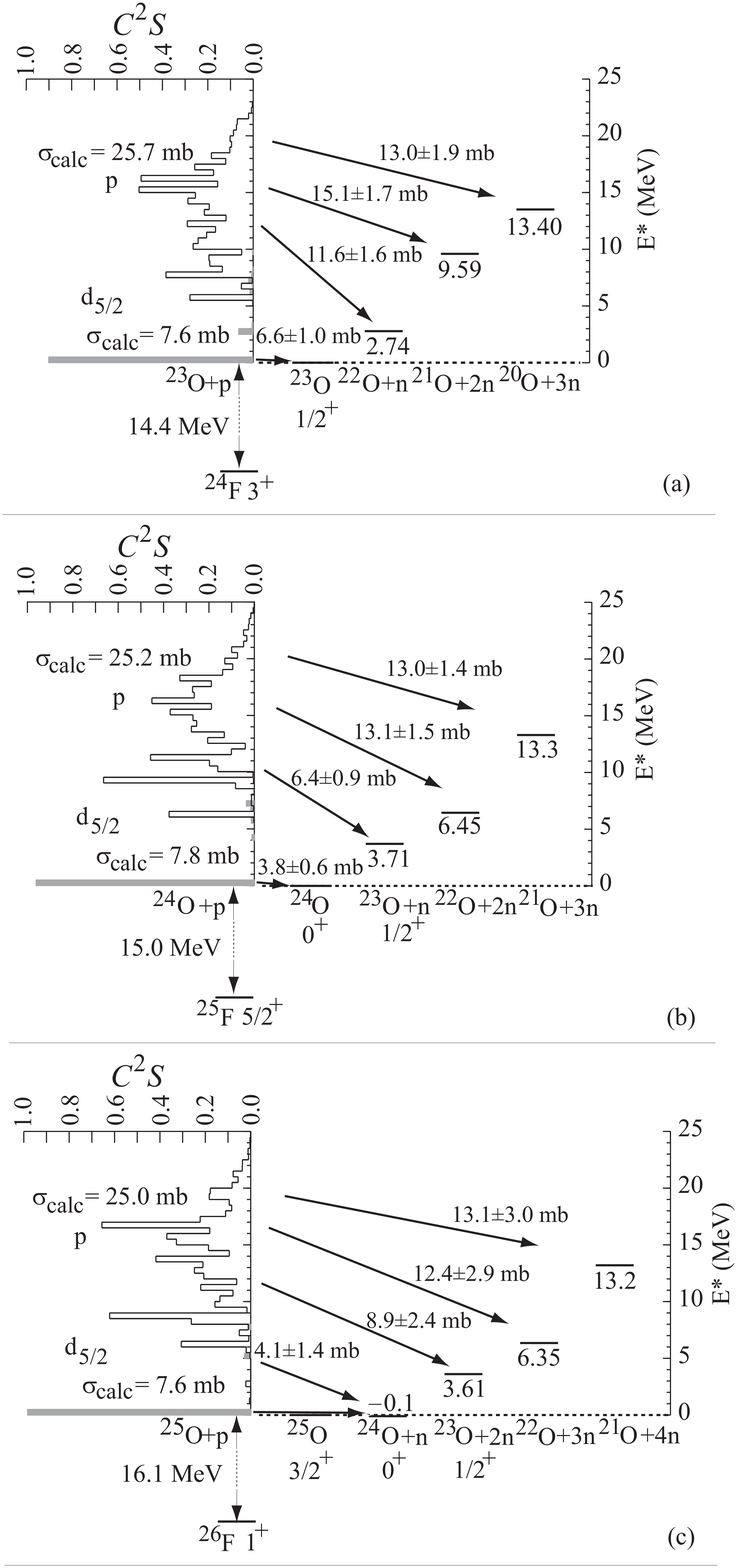}
 \caption{Decay schemes for $^{24}$F (a), $^{25}$F (b), and $^{26}$F. The calculated excitation energy distribution of the spectroscopic factors ($C^2S$/500keV) for $d_{5/2}$ (shaded bars) and $p$-shell protons are indicated on the left. The cross sections in the Figure are calculated as the product of
the spectroscopic factors and the single-particle cross sections of Figure \ref{f:excitation}. }
  \label{f:schemes}
\end{figure}

For the population of the ground state of $^{25}$O in the proton knock-out from $^{26}$F only an upper limit can be extracted because $^{25}$O is unbound and decays to $^{24}$O. The ground state of $^{24}$O can also be populated from the decay of excited states in $^{25}$O. Thus, the measured cross section of 4.1$\pm1.4$ mb correspond to a sum of the contributions from the decay of the ground-state and excited states of $^{25}$O (see also Figure \ref{f:schemes}(c) and the following section).

\subsection{Population of Other Oxygen Isotopes}

\begin{figure}
  \includegraphics[width=0.8\linewidth]{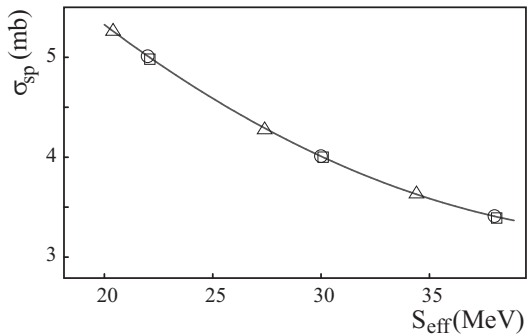}
  \caption{Single proton knock-out cross sections as a function of the effective proton separation energy. Single-particle cross sections for proton removal from the $p$-shell have been calculated for three excitation energies in $^{24}$F (triangles), $^{25}$F (circles), and $^{26}$F (squares). The solid line shows a fit that was used to calculates the toal $p$-wave removal cross section.}
  \label{f:excitation}
\end{figure}

The cross section for the population of other oxygen isotopes in these breakup reactions can only be estimated because several production mechanisms can contribute. In addition to the single-proton knock-out followed by subsequent neutron emission, direct population of the ground state as well as bound excited states have to be taken into account. 

Within the eikonal approach one can calculate the single-particle distribution due to contributions from the knock-out of protons from the $p$-shell. The spectroscopic strength distribution as calculated within the shell model using the universal $sd$ (USD) interaction \cite{Bro88} is shown on the left side of Figure \ref{f:schemes}. The single-particle cross section calculated in the eikonal model depends on the effective proton separation energy which is the sum of the proton separation energy and the excitation energy of the specific state. The proton separation energies for $^{24}$F, $^{25}$F, and $^{26}$F are 14.4 MeV, 15.0 MeV, and 16.1 MeV, respectively, Figure \ref{f:excitation} shows that the proton-removal cross sections are very similar for the three fluorine isotopes. In order to extract the total proton-removal cross sections from the $p$-shell the spectroscopic distributions were folded with the excitation-energy dependent stripping cross section of Figure \ref{f:excitation}. The values for the $p$-shell 
proton-removal cross sections are 25.7 mb, 25.2 mb and 25.0 mb for $^{24}$F, $^{25}$F, and $^{26}$F, respectively, and are listed (p,$\sigma_{calc}$) in Figure \ref{f:schemes}. 

These populations are located above the one-, two-, three-, and even the four-neutron evaporation channel. The decay branches to the individual neutron channels depend on the details of the excited states in these nuclei. As mentioned above other processes also contribute to the population of these channels, so that calculating these decay branches is not useful. The only quantitative statement that can be extracted from the present data is that the calculated $p$-shell proton-removal cross section should not exceed the summed cross section of the open neutron channels.

For the reaction ($^{24}$F,$^{23}$O) the 1-n to 3-n channels have to be taken into account. The threshold for the 4-n channel ($^{19}$O) at 21 MeV is located above the population for the removal of protons from the $p$-shell. 
The sum of the cross sections for these three neutron decay channels (population of $^{22-20}$O) as listed in Table \ref{t:cross} is 39.7$\pm$3.0~mb, which has to be compared to a total proton-removal cross section from the $p$-shell of 25.7~mb. These cross sections are also indicated in Figure \ref{f:schemes}.

The sum of the population of the 1-n to 3-n channel ($^{23-21}$O) in the reaction ($^{25}$F,$^{24}$O) is 32.4$\pm$2.2 mb. This value is only slightly larger than the calculated proton-removal cross section of 25.2 mb. However, the threshold for the 4-n channel ($^{20}$O) is located at 17.1 MeV, which is below the highest energies of the distribution of $p$-shell proton removal so that it also could contribute.

The reaction ($^{26}$F,$^{25}$O) has to be treated differently because $^{25}$O is unstable and thus the contribution from $p$-shell removal can not be distinguished from the removal of the $d_{5/2}$ proton. The total calculated single proton-removal cross section from $d$- and $p$-wave knock-out is 32.6 mb. The open channels as indicated in Figure \ref{f:schemes} are the 1-n through 4-n decay, which corresponds to a cross section of 38.5$\pm$5.0 mb. It might be even possible that the 5-n channel ($^{20}$O) with a separation energy of 17 MeV can contribute.

For all three reactions studied, the predicted $p$-shell removal cross sections are consistent with the measured populations of the open neutron decay channels.

It is interesting to note that the widely used empirical fragmentation model 
 EPAX \cite{Sum90,Sum00} yields cross sections of the order of 10$-$20 mb for the population of these oxygen isotopes in agreement with the values listed in Table \ref{t:cross}.

The recent measurements of fragmentation of unstable neutron-rich oxygen isotopes also included single proton knock-out cross sections \cite{Lei02}. These measurements were performed at significantly higher energies ($>$500 MeV/nucleon) and yielded cross sections of $\sim$20 mb. These measurements were not performed as close to the dripline as the current experiment and the populated nitrogen isotopes all had bound excited states, so a direct comparison to the present analysis is not possible. Only the reaction of the most neutron-rich projectile $^{21}$O populates a nucleus with no known bound excited states ($^{20}$N). Unfortunately no cross section is quoted for this reaction \cite{Lei02}.

%*************************************************************************************
\section{Conclusions}

The results of the single-proton knock-out reactions were analyzed within the a theoretical approach combining many-body shell model and eikonal theory. 
The cross section for the reaction ($^{24}$F,$^{23}$O) is consistent with a single particle description of these nuclei, while the cross sections for the reaction ($^{25}$F,$^{24}$O) shows a suppression of $\sim$50\%. However, experiments with higher precision are necessary in order to infer a change of the structure.
The knock-out reactions on $^{24}$F and $^{25}$F are special because of the non-existence of bound excited states in $^{23}$O and $^{24}$O, respectively. The present method promises to be a powerful tool to extract spectroscopic information for nuclei at the very edge of the driplines because they typically do not have any bound excited states. Other nuclei will require more sophisticated experimental arrangements, including $\gamma$-ray coincidences.

\section{Acknowledgments}
\label{s:thanks}
This work has been supported by the National
Science Foundation grant number PHY01-10253 and the PHY00-70911.
%*************************************************************************************

%*************************************************************************************
\end{document}